\documentclass{article}
\usepackage{spconf,amsmath,graphicx,hyperref}
\usepackage{cite}
\usepackage{indentfirst}
\usepackage{amssymb}
\usepackage{algorithm}
\usepackage{algorithmic}
\usepackage{booktabs}
\usepackage{multirow}
\usepackage{xurl}
\makeatletter
\g@addto@macro{\UrlBreaks}{\do\a\do\b\do\c\do\d\do\e\do\f\do\g\do\h\do\i\do\j
\do\k\do\l\do\m\do\n\do\o\do\p\do\q\do\r\do\s\do\t\do\u\do\v\do\w\do\x\do\y\do\z
\do\A\do\B\do\C\do\D\do\E\do\F\do\G\do\H\do\I\do\J
\do\K\do\L\do\M\do\N\do\O\do\P\do\Q\do\R\do\S\do\T\do\U\do\V\do\W\do\X\do\Y\do\Z
\do0\do1\do2\do3\do4\do5\do6\do7\do8\do9}
\makeatother
\usepackage{hyperref}
\def\x{{\mathbf x}}
\def\L{{\cal L}}

\title{Emotion-Aware Speech Generation with Character-Specific Voices for Comics}
\name{Zhiwen Qian, Jinhua Liang, Huan Zhang}
\address{School of Electronic Engineering and Computer Science\\ 
Queen Mary University of London}

\begin{document}
\ninept
\maketitle
\begin{abstract}
This paper presents an end-to-end pipeline for generating character-specific, emotion-aware speech from comics. The proposed system takes full comic volumes as input and produces speech aligned with each character’s dialogue and emotional state. An image processing module performs character detection, text recognition, and emotion intensity recognition. A large language model (LLM) performs dialogue attribution and emotion analysis by integrating visual information with the evolving plot context. Speech is synthesized through a text-to-speech (TTS) model with distinct voice profiles tailored to each character and emotion. This work enables automated voiceover generation for comics, offering a step toward interactive and immersive comic reading experience.
\end{abstract}
\section{Introduction}
\label{sec:intro}

Digital comics have become increasingly popular, offering opportunities to engage readers through intelligent media processing. A promising direction is automatic audiobook generation with comics, which aims to convert static dialogue into dynamic and expressive speech. Such systems could enrich digital reading experiences by making comics more immersive and accessible.

A comics speech generation pipeline typically requires: (1) image analysis for detecting characters and text regions, (2) character identification within panels, (3) speaker attribution to link text to speakers, (4) emotion recognition for dialogue tone, and (5) text-to-speech (TTS) synthesis with emotion control. Each of these components has been studied separately: structural analysis and scene graph generation (SGG) \cite{johnson2015image,krishna2017visual,xu2017scene,zellers2018neural,faster_r_cnn}, OCR for text extraction \cite{smith2007overview,shi2017crnn,mangaocr2023}, speaker attribution via layout and contextual features \cite{li2024manga109dialog,li2024zeroshot}, emotion annotation resources such as KangaiSet \cite{theodose2023kangaiset}, and recent emotion-controllable audio generative systems \cite{hisariya2025bridging,liang2025audiomorphix,jia2018transfer,cho2024emosphere}. However, integrating them into a complete pipeline remains underexplored.

Our work builds on major comic datasets, particularly Manga109 \cite{matsui2017manga109} and its derivatives: Manga109Dialogue, which links dialogues to speaker IDs \cite{li2024manga109dialog}, and KangaiSet, which provides character facial emotion labels \cite{theodose2023kangaiset}. These resources enable tasks such as dialogue attribution, character recognition, and emotion classification. However, several challenges remain underexplored: (i) reliable visual identification of characters is difficult due to the long-tail distribution of appearances, (ii) speaker attribution is non-trivial when dialogue text is located far from the speaker in the comic, and (iii) emotion classification suffers from imbalanced datasets with rare classes.

Inspired by agent-based approaches~\cite{liu2025wavjourney,liang2024wavcraft}, we address these challenges by combining vision-based approches and large language models (LLMs). CNN-based approaches \cite{lecun1998lenet,krizhevsky2012alexnet,simonyan2015vgg,he2016resnet,jia2024bridging} remain effective for character and emotion recognition, but require labeled data. Clustering-based unsupervised methods \cite{zhang2022unsupervisedmangacharacterreidentification,soykan2023identityaware} mitigate annotation costs by discovering recurring identities. Meanwhile, LLMs such as ChatGPT~\cite{openai2024chatgpt} provide strong contextual reasoning, enabling multimodal integration of dialogue content and spatial layout for speaker and emotion inference. Recent zero-shot methods \cite{li2024zeroshot} highlight the potential of combining visual cues with textual reasoning for robust speaker assignment.

Among existing works, \cite{li2024manga109dialog} and \cite{li2024zeroshot} are the most closely related to our study. The former emphasizes speaker attribution by linking dialogue bubbles to character identities, while the latter explores LLM-based reasoning for character recognition under zero-shot conditions. In contrast, our work focuses on both speaker and emotion prediction, and further integrates these components into a complete end-to-end system for comic audiolization.  Example generated audio-comics can be found in our demo page~\url{www.notion.so/Demo-Page-26fd763d742080f0be6ff7c522d7ea23}. Our contributions are as follows: 
\begin{itemize}
    \item We design a comprehensive pipeline for comic speech generation.
    \item We propose a novel emotion prediction method that simplifies classification on images followed by LLM-based fine classification.
    \item We systematically investigate the capability of LLM to leverage both visual and textual information for predicting speakers and emotions from comic dialogues.
\end{itemize}

\begin{figure*}[htbp]
    \centering
    \includegraphics[width=0.95\textwidth]{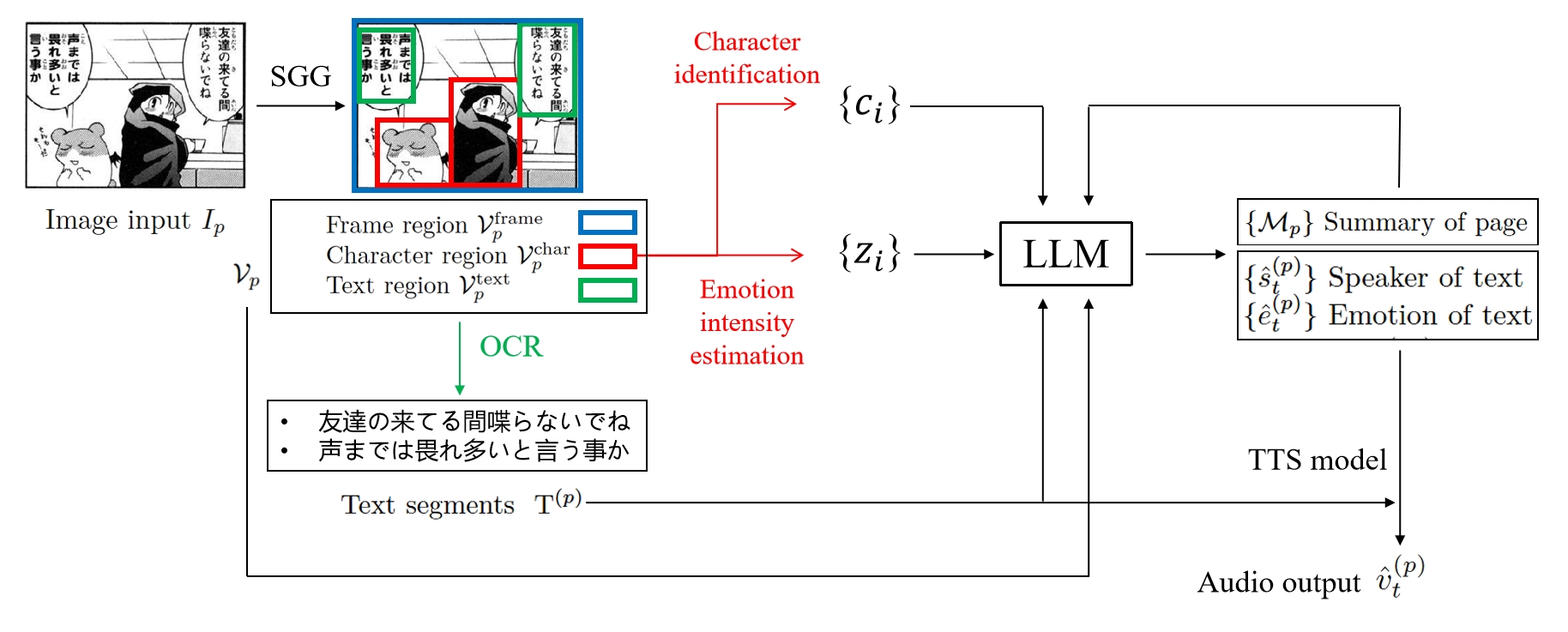}
    \caption{Overall pipeline of the proposed system for emotion-aware, character-specific speech generation from comics.}
    \label{fig:overall-pipeline}
\end{figure*}

\section{Methodology}

\subsection{Problem Setting}
To generate emotionally expressive, character-specific speech for each page of a comic, we propose the overall pipeline of our system as shown in Figure~\ref{fig:overall-pipeline}. The input consists of a set of page images $\{I_p\}$, along with a character reference dataset $\mathcal{R}$. These samples serve as supervision for training the character identification network.

Each page $I_p$ contains a set of text segments ${T^{p}}$, where each $\{T^{p}\}$ is extracted from detected text regions using an OCR model $f_{\text{ocr}}$. The visual elements $\mathcal{V}_p$ are extracted via scene graph generation. Let $\mathcal{M}_{p-1} = (\mathcal{G}_{p-1}, \mathcal{L}_{p-1})$ denote the LLM memory state before page $p$, where $\mathcal{G}_{p-1}$ is the accumulated \textit{global plot summary} up to page $p-1$, and $\mathcal{L}_{p-1}$ is the \textit{local summary} of page $p-1$. The LLM predicts the speaker identity $\hat{s}_{t}^{p}$ for text segment $t$ and the corresponding emotion label $\hat{e}_{t}^{p}$ on page $p$ as follows:
\begin{equation}
\{\hat{s}_{t}^{p}, \hat{e}_{t}^{p}\}_{t=1}^{N^{\text{text}}_p} = \mathrm{LLM}(\mathcal{V}_p, \{T^{p}\},\{\hat{c}_i\}, \{z_i\}, \mathcal{G}_{p-1}, \mathcal{L}_{p-1}),
\end{equation}
where $\{\hat{c}_i\}$ and $\{z_i\}$ are the results of character identification and estimation of emotion intensity, and $N^{\text{text}}_p$ is the number of text segments on page $p$.

The final audio output for each dialogue is synthesized using a reference-conditioned TTS model:
\begin{equation}
    \hat{v}_{t}^{p} = \mathrm{TTS}(T^{p} \mid \hat{s}_{t}^{p},~\hat{e}_{t}^{p}).
\end{equation}

\subsection{Scene Graph Generation}

We apply a Faster R-CNN model \cite{faster_r_cnn} $f_{\text{SGG}}$ to each page image $I_p$ to extract:
\[
\mathcal{V}_p = \{\mathcal{V}_p^{\text{frame}},~ \mathcal{V}_p^{\text{text}},~ \mathcal{V}_p^{\text{char}}\} = f_{\text{SGG}}(I_p) 
\]

Frames are further ordered using a reading-order algorithm~\cite{mtap_matsui_2017}, producing a sequence $\mathcal{F}_p = [f_1, f_2, \dots, f_n]$, which is later used in prompt construction.

\subsection{Character Identification}
For each comic, we train a separate character identification model $f_{\text{char}}$ using a small set of annotated reference images $R = \{R_1, R_2, \dots, R_k\}$, where each $R_i = \{(x_{i,j}, y_i)\}_{j=1}^{n_i}$ consists of $n_i$ image crops labeled as character $i$.

To enhance robustness, we construct an extended training set $R' = R \cup R_{\text{others}}$, where $R$ consist of $k$ different categories while $R_{\text{others}}$ contains negative examples randomly sampled from characters in other comics and assigned to a unified ``others'' class. This transforms the task into a $(k + 1)$-way classification problem.

We adopt a ResNet-50 backbone pretrained on ImageNet, fine-tuned on $R'$ with standard data augmentation techniques such as rotation and scaling. The trained classifier $f_{\text{char}}$ is used to assign each detected character instance $x$ to one of the $k$ main characters or the ``others'' class:
\[
\hat{y} \leftarrow f_{\text{char}}(x),~\hat{y} \in \{1, \dots, k+1\}.
\]

We then define $F_{\text{char}}(\mathcal{V}_p^{\text{char}})$ as the batched output that returns predicted identities for all character instances on page $p$:
\[
\{\hat{c}_i\}_{i=1}^{N^{\text{char}}_p} = F_{\text{char}}(\mathcal{V}_p^{\text{char}})
\]

where $N^{\text{char}}_p$ denotes the number of character instances.
\subsection{Emotion Intensity Estimation}

The KangaiSet dataset defines seven emotion categories for comic character faces. However, the distribution is highly imbalanced. For example, \textit{neutral} accounts for 35.02\% of the samples, while \textit{disgust} represent only 0.47\%. This makes multi-class classification biased. To address this, we reformulate the task as binary emotion intensity classification:
\[
e \leftarrow f_{\text{emo}}(x),~e \in \{0,~1\},
\]
where $e$ denotes whether the input character instance exhibits visible emotional expression. During inference, instead of using the binary output directly, we pass the raw logit $z_i \in \mathbb{R}$ as an emotion strength score.

The batched version is defined as:
\[
\{z_i\}_{i=1}^{N^{\text{char}}_p} = F_{\text{emo}}(\mathcal{V}_p^{\text{char}}).
\]
These scores $\{z_i\}$ are used as auxiliary inputs to the LLM to predict emotion categories for dialogue segments, leveraging the model's contextual reasoning capabilities.

\begin{table*}[t]
\centering
\caption{Speaker and Emotion Prediction Percentage Across Methods}
\label{tab:llm-speaker}
\begin{tabular}{lccc|cc}
\toprule
\multirow{2}{*}{\textbf{Method}} 
& \multicolumn{3}{c|}{\textbf{Speaker Accuracy (\%)}} 
& \multicolumn{2}{c}{\textbf{Emotion Accuracy (\%)}} \\
 & Easy & Hard & Total & Neutral & Total \\
\midrule
Rule-based (short dist.)  & 71.4 & 22.7 & 63.4 & --- & --- \\
Rule-based (frame dist.)  & 81.6 & 22.1 & 71.5 & --- & --- \\
Manga109Speaker (best)         & \textbf{84.8} & 30.7 & \textbf{75.7} & --- & --- \\
Zero-Shot Multimodal (iter2)         & 52.4 & \textbf{51.3} & 51.8 & --- & --- \\
\midrule
Ours Setting A (GT char only)       & 77.5 & 19.4 & 63.6 & \textbf{45.7} & 39.6 \\
Ours Setting B (GT char + emo)      & \textbf{81.7} & \textbf{21.8} & \textbf{66.4} & 35.8 & 40.8 \\
Ours Setting C (Pred char + emo)    & 79.2 & 20.5 & 64.8 & 34.6 & \textbf{41.2} \\
\bottomrule
\end{tabular}
\end{table*}

\section{Experiment and evaluation}
\subsection{Dataset Setup}
Since our task requires associating textual emotion with spoken dialogue, we link KangaiSet and Manga109Dialogue through shared character instances. This creates a labeled subset of emotion-tagged dialogue lines. 

To ensure reliable evaluation, we select the 20 comic titles in KangaiSet with the largest number of emotion annotations as our test set, which is excluded entirely from the training process. The remaining titles are used for training and development.

\subsection{Image Processing}
The SGG model is trained on the training portion of Manga109 using annotated bounding boxes for character bodies, text regions, and frames. 

Dialogue text is extracted using manga-ocr\cite{mangaocr2023}, a high-accuracy OCR system optimized for Japanese comic content. 

Detected frames are post-processed using the method described in\cite{mtap_matsui_2017} to determine reading order and frame merging. During prompt construction for the LLM, the system encodes all text and characters contained within each frame in reading order, followed by any text or characters not assigned to any frame.

By default, the system assumes that a comic image contains two pages, and applies a two-column layout splitting algorithm, except for four-panel manga, which are split into four vertical strips. Exceptions to these layout assumptions may lead to incorrect frame sequencing during reading order determination.

\textbf{Character Recognition (CR):} We identify the main characters in each test-set comic as those appearing more than 50 times. For each character, we select 40 annotated images and combine them with 40 other class samples (from different comics) to train a ResNet-50 classifier. Standard data augmentation techniques such as random rotation and scaling are applied during training.

We evaluate character identification performance across 20 comic titles, achieving an average top-1 accuracy of 62.9\%. This demonstrates moderate success given the constraint of limited reference samples. Performance in most comics ranged from 55\% to 75\%, with variation attributable to factors such as consistency in character design, frequency of appearance, and visual similarities between characters.

These results highlight both the potential and current limitations of per-title character-specific classifiers trained on small-scale reference sets. Errors in this stage also propagate into downstream tasks.

\textbf{Emotion Intensity (EI):} the binary classifier is trained using the simplified version of KangaiSet, where all \textit{non-neutral} emotions are grouped into a single class representing strong emotion. A pretrained ResNet-50 is fine-tuned on the training set, and the model's logit output is used as the estimated emotion intensity score for each character.

\begin{table}[htbp]
\caption{Hyper-parameters of different modules during training.}
\centering
\begin{tabular}{lccc}
\hline
\textbf{Parameter} & \textbf{SGG} & \textbf{CR} & \textbf{EI} \\
\hline
Learning rate & 1e-3  & 1e-4 & 1e-4 \\
Batch size    & 2     & 8     & 32 \\
Optimizer     & SGD   & SGD  & SGD \\
Epochs        & 20     & 40    & 20 \\
\hline
\end{tabular}
\label{tab:training_params}
\end{table}

The model achieves an overall accuracy of 70.9\%, with asymmetric performance across the two classes: 87.8\% accuracy for strong expressions versus only 41.6\% for \textit{neutral} ones, possibly because neutral expressions are often hard to distinguish from weak emotions. 

Despite attempts to rebalance the training set through oversampling and loss reweighting techniques, the gap between classes remained significant. This suggests that neutral expressions are often hard to distinguish from weak emotions. Additionally, KangaiSet was labeled emotion based solely on facial appearance without considering the surrounding context or narrative flow, further increasing the labeling noise for neutral samples.

Table~\ref{tab:training_params} shows the training parameters of the three image processing models. The pretrained parameters are taken from the default torchvision models.

\subsection{Speaker Prediction}
In terms of LLM model, we adopted GPT-4, which demonstrated strong consistency in structured output generation across pages. Notably, this is also the model used in the zero-shot baseline proposed by \cite{li2024zeroshot}, ensuring a fair comparison. For each page, the model was prompted with detected visual elements, as well as dialogue text and narrative summaries. The model returned the predicted speaker, emotion label, and updated plot summaries, enabling seamless integration into the full pipeline. Table~\ref{tab:llm-speaker} shows the accuracy of speaker and emotion prediction among different methods.

\textbf{Baseline: }We compare our method against several baselines. The first group is derived from the Manga109Speaker\cite{li2024manga109dialog} study, including two rule-based heuristics. In addition, we include their best-performing deep learning model. The other is taken from \cite{li2024zeroshot}, which introduces a zero-shot, multimodal reasoning framework for speaker attribution. We report the result from their Multimodal Iteration 2 setting, as it achieves the best overall performance among all configurations in their study.

We evaluate three \textbf{model variants} with increasing realism: (A) gold character identity only, (B) gold identity with predicted emotion intensity, and (C) predicted identity with predicted emotion intensity. For efficiency, our evaluation was conducted on the first 10 images of each of the 20 test-set comic titles.

\begin{table}[htbp]
\centering
\caption{5-way emotion recognition in Setting C with classification performance (\%) and number of support samples.}
\label{tab:emotion-report}

\begin{tabular}{lcccc}
\toprule
\textbf{Label} & \textbf{Precision} & \textbf{Recall} & \textbf{F1} & \textbf{\#Support} \\
\midrule
Neutral    & 56.6 & 34.6 & 42.6 & 159 \\
Surprise   & 15.7 & 52.6 & 24.2 & 38 \\
Anger      & 42.2 & 47.9 & 44.9 & 73 \\
Happiness  & 76.8 & 39.9 & 52.5 & 158 \\
Sadness    & 47.1 & 50.0 & 48.5 & 48 \\
\midrule
Micro avg  & 44.6 & 41.4 & 42.9 & 476 \\
Macro avg  & 47.5 & 45.0 & 42.9 & 476 \\
Weighted avg & 56.5 & 41.4 & 45.4 & 476 \\
\bottomrule
\end{tabular}
\end{table}

\textbf{Compared to the baselines}, our Setting C achieves higher accuracy than the Zero-Shot model from on both easy and total speaker attribution. However, it underperforms relative to the Manga109Speaker deep learning baseline, particularly in the hard cases. This may be because, when the dialogue text and the speaker are not located within the same frame, the deep learning model can learn implicit spatial patterns to resolve speaker associations. In contrast, the LLM relies primarily on textual layout and proximity heuristics, making its behavior more similar to rule-based methods under these challenging conditions. Our Setting C also slightly trails the rule-based frame-distance heuristic on total accuracy. These results suggest that LLM-based reasoning still struggles in hard cases, reflecting a key limitation: while humans can infer visual context beyond text, the LLM lacks sufficient scene understanding even when given character identity labels.

\textbf{Impact of Incorrect Character Identity Input:}
In the transition from B to C, the speaker prediction accuracy drops by approximately 2\% across all difficulty levels. This indicates that incorrect character identity predictions can mislead the LLM when attributing dialogue. On the emotion side, neutral class accuracy again drops, while overall accuracy slightly improve, potentially due to variance in how LLMs interpret text cues rather than visual features.

\subsection{Emotion Prediction}
\textbf{Effectiveness of Emotion Intensity Input:}
Comparing Settings A and B, we observe a slight improvement of 1.2\% in overall emotion classification accuracy when emotion intensity is included. However, accuracy on the neutral class decreases significantly. This aligns with our emotion intensity model's behavior: neutral expressions are often over-predicted as strong, which likely introduces noise into the LLM's emotion reasoning.

\begin{figure}[htbp]
\centering
\includegraphics[width=0.90\linewidth]{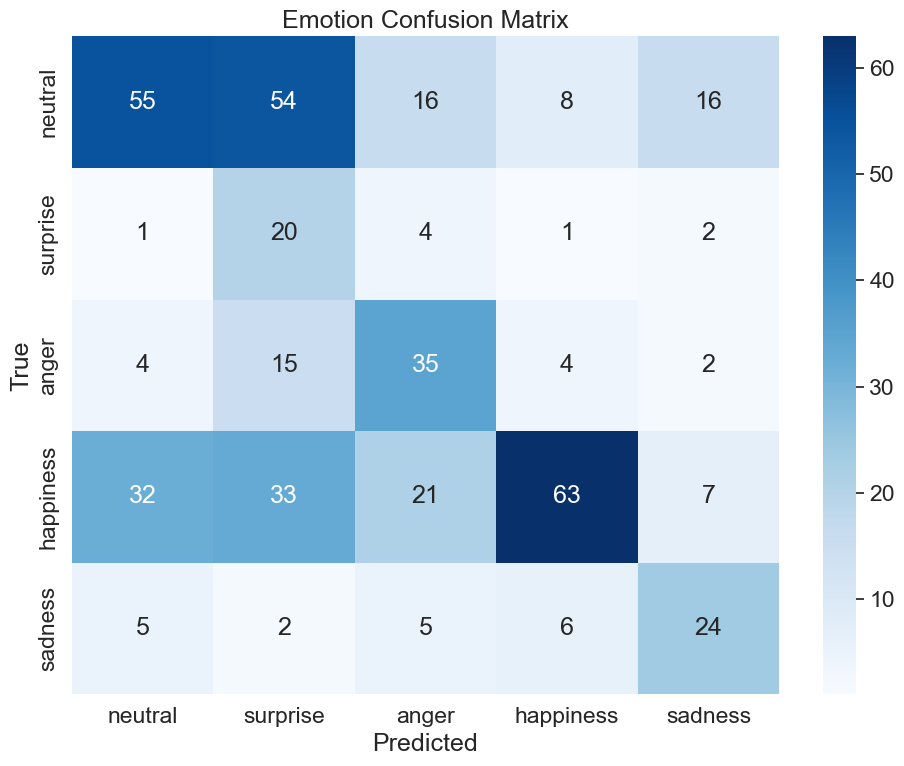}
\caption{Confusion matrix for emotion classification in Setting C.}
\label{fig:emo-conf}
\end{figure}

Table~\ref{tab:emotion-report} presents the classification report for 5-way emotion recognition in Setting C. The emotion categories disgust and fear were excluded due to their extremely low sample counts in the dataset. Figure~\ref{fig:emo-conf} further illustrates the prediction distribution through a confusion matrix.

Among the remaining classes, surprise shows relatively high recall (52.6\%) but very low precision (15.7\%), indicating a large number of false positives. Upon manual inspection, we found that the LLM tends to label interrogative sentences as surprise. As a result, many predictions of surprise originate from sentences that are labeled as neutral in the ground truth. However, from the perspective of voice generation, such misclassifications may not result in perceptually inappropriate expression.

While the overall emotion classification accuracy of the LLM appears low (under 42\%), we conducted a manual inspection to better understand this result. In comics, the emotional meaning of a character's face is often ambiguous or open to multiple interpretations. This observation indicates that current accuracy metrics may underestimate the true quality of emotion inference in our system, especially given the inherent subjectivity and visual ambiguity of comic-style emotional expression. A manual inspection example can be found in our demo page. 

The neutral class, which constitutes a large portion of the dataset, remains difficult to predict accurately. Its low recall (34.6\%) is consistent with previous observations that subtle or ambiguous expressions are frequently misclassified.

\subsection{Joint Accuracy for Speaker and Emotion}
The ultimate goal of our system is to generate speech that reflects both the correct speaker identity and the appropriate emotional tone for each dialogue segment. To evaluate this end-to-end goal, we compute the proportion of dialogue segments in Setting~C where both the predicted speaker and emotion label are correct. This joint accuracy was found to be \textbf{20.4\%} across the evaluated pages.

While this value appears modest, it reflects the compounded difficulty of two challenging tasks, subject to upstream errors and inherent ambiguity. Improving joint performance remains a key direction for future work, particularly through better integration of visual context and multi-turn narrative modeling.

\section{Conclusion}
This paper proposed an end-to-end pipeline for generating character-specific, emotion-aware speech from comics. The system combines vision-based character and emotion recognition with LLM-based dialogue attribution and emotion inference, followed by reference-conditioned TTS for expressive voice synthesis.

Experimental results demonstrate that the system can achieve moderate accuracy in character recognition and emotion classification, despite the limited availability of annotated data. The character identification module performs reliably across diverse comic styles, but errors in this stage propagate downstream and impact speaker attribution. Our LLM-based dialogue and emotion inference show potential but are sensitive to the complexity of visual input and subject to ambiguity in emotion labels, particularly for the neutral class. Human inspection reveals that some errors can still produce perceptually appropriate outputs, highlighting the gap between evaluation metrics and human judgment.

Overall, our work illustrates the feasibility of automatic, expressive comic voiceover and lays the groundwork for more robust and human-aligned audiovisual storytelling systems.

% Below is an example of how to insert images. Delete the ``\vspace'' line,
% uncomment the preceding line ``\centerline...'' and replace ``imageX.ps''
% with a suitable PostScript file name.
% -------------------------------------------------------------------------

%\vfill
%\pagebreak

\bibliographystyle{IEEEbib}
\bibliography{refs}

\end{document}